\documentstyle[12pt,epsf]{article}
\def\draftversion{N}			    % Y for draft, N for final version
\def\note[#1]#2{\message{(#1)}\if\draftversion Y{\noindent\em #2}\fi}
\newif\ifenglish\englishfalse               % or American
\if \draftversion Y
\fi
\font\bbxii=msbm10 scaled 1200
\font\bbx=msbm10
\font\bbvii=msbm7
\newfam\bbfam
\def\bb{\fam\bbfam\bbxii}
\textfont\bbfam=\bbxii
\scriptfont\bbfam=\bbx
\scriptscriptfont\bbfam=\bbvii
\def\twovector[#1,#2]{\left(\begin{array}{c} #1 \\ #2 \end{array}\right)}
\def\twomatrix[#1,#2;#3,#4]%
  {\left(\begin{array}{cc}
    #1 & #2 \\
    #3 & #4
  \end{array}\right)}
\def\half{{\scriptstyle{1\over2}}}  % One half
\def\quarter{{\scriptstyle{1\over4}}} % One quarter
\def\Z{{\bb Z}}			    % Set of integers
\def\Re{\mathop{\rm Re}}	    % Real Part
\def\Im{\mathop{\rm Im}}	    % Imaginary Part
\def\dt{\delta\tau}		    % MD integration step size
\def\trjlen{\tau}		    % MD trajectory length
\def\pacc{P_{\hbox{\rm\tiny acc}}}  % Acceptance probability
\def\identity{{\bb I}}		    % Identity matrix
\def\dH{{\delta H}}		    % Energy change over trajectory
\def\erfc{\mathop{\rm erfc}}	    % Complementary error function
\def\Dd{\Delta\delta}		    % Difference between forward and backward
\def\half{{\scriptstyle{1\over2}}}  % One half
\def\quarter{{\scriptstyle{1\over4}}} % One quarter
\begin{document}

\title{Instabilities and Non-Reversibility of Molecular Dynamics Trajectories}

\author{R.~G.~Edwards, Ivan Horv\'ath, and A.~D.~Kennedy \\[1ex]
  Supercomputer Computations Research Institute, \\
  Florida State University, Tallahassee, Florida 32306-4052, U.S.A.}
\date{FSU--SCRI--96--49\\[0.75ex] June 13, 1996}

\maketitle

\begin{abstract}
  \noindent The theoretical justification of the Hybrid Monte Carlo algorithm
  depends upon the molecular dynamics trajectories within it being exactly
  reversible. If computations were carried out with exact arithmetic then it
  would be easy to ensure such reversibility, but the use of approximate
  floating point arithmetic inevitably introduces violations of reversibility.
  In the absence of evidence to the contrary, we are usually prepared to accept
  that such rounding errors can be made small enough to be innocuous, but in
  certain circumstances they are exponentially amplified and lead to blatantly
  erroneous results. We show that there are two types of instability of the
  molecular dynamics trajectories which lead to this behavior, instabilities
  due to insufficiently accurate numerical integration of Hamilton's equations,
  and intrinsic chaos in the underlying continuous fictitious time equations of
  motion themselves. We analyze the former for free field theory, and show that
  it is essentially a finite volume effect. For the latter we propose a
  hypothesis as to how the Liapunov exponent describing the chaotic behavior of
  the fictitious time equations of motion for an asymptotically free quantum
  field theory behaves as the system is taken to its continuum limit, and
  explain why this means that instabilities in molecular dynamics trajectories
  are not a significant problem for Hybrid Monte Carlo computations. We present
  data for pure $SU(3)$ gauge theory and for QCD with dynamical fermions on
  small lattices to illustrate and confirm some of our results.
  {\parfillskip=0pt\par}
\end{abstract}

\section{Introduction}

The goal of this paper is to study the effect of errors introduced by working
with finite-precision arithmetic on Hybrid Monte Carlo (HMC) and related
algorithms.

The HMC algorithm \cite{kennedy87a,duane87a,kennedy90a} allows us to generate an
ensemble of field configurations which are selected from some probability
distribution. The algorithm is ``exact,'' in the sense that it is a Markov
process which converges to the desired distribution with no systematic errors
provided that the fictitious time molecular dynamics (MD) trajectories within it
are exactly reversible and area preserving, that the computation of the action
for the Metropolis accept/reject step is exact, and that we have a supply of
perfectly random numbers. It does not require that the MD integration or any
conjugate gradient (CG) ``inversion'' of the fermionic kernel be carried out
exactly\footnote{The initial CG vector must be chosen in a time-symmetric way.}.
Indeed, if we were to carry out all computations using exact arithmetic (i.e.,
with no rounding errors whatsoever) then the leapfrog integration scheme
provides a method of integrating Hamilton's equations while maintaining exact
reversibility and area preservation.

All numerical computations carried out using floating point arithmetic are
subject to rounding errors, but unless these errors are amplified exponentially
we do not normally consider them to be a serious problem. The reason for this is
that if a sensible rounding mode\footnote{If we were to truncate all floating
point numbers towards zero after each operation then the error after $N$
operations would be $\varepsilon N$, and the cost would grow twice as rapidly
with $N$ --- but still logarithmically.} is used for the elementary floating
point operations then the error after a sequence of $N$ operations, each of
which gives rise to an error of magnitude $\varepsilon$, is $\varepsilon\sqrt
N$. The cost of working with $d$ fractional digits grows approximately
linearly\footnote{On a typical computer chip doubling the number of digits
requires doubling the number of gates in the adder and multiplier and increasing
the clock period to allow for the worst-case carry propagation. Thus
asymptotically the number of gates, the clock period, and the amount of memory
would increase linearly with the number of digits~$d$. In practice it is better
to reduce the carry propagation time to grow as $\ln d$ at the cost of
increasing the number of gates as $d\ln d$. This leads to an asymptotic cost
which grows as $d(\ln d)^2$.} with $d$ and gives $\varepsilon=10^{-d}$.
Therefore if we require an answer with an error smaller in magnitude than
$\delta$ in magnitude the cost is proportional to $-\log_{10}(\delta/\sqrt N)$,
which grows only like $\ln N$. This is a very small correction\footnote{For real
computers the number of digits available is often ``quantized'' in units of
words, and increasing the precision from about $7$ digits (IEEE single
precision) to about $14$ digits (IEEE double precision) may be large, so
asymptotic estimates must be treated with caution.} to the growth of the cost of
the HMC algorithm as the volume and correlation length of the system is
increased.

All the arguments we present as to the conditions under which reversibility is
maintained apply to area preservation as well, but we shall concentrate on the
former as the latter is less easy to determine empirically.

\section{A Brief Survey of Algorithms}

Let us very briefly summarize the various advantages and disadvantages of
several variants of the HMC algorithm for dynamical fermion Quantum
Chromodynamics (QCD) computations.

\subsection{Hybrid Monte Carlo (HMC)}

The algorithm is exact even if only a cheap approximate CG solution is carried
out for computing the force along a MD trajectory, and such a CG solution is
needed once per integration step. Nevertheless, an accurate CG solution is
needed for the Metropolis accept/reject test once per trajectory.

If we believe that the behavior of free field theory carries over to
interacting theories like QCD, then choosing trajectory lengths which grow
linearly with the correlation length of the system $\trjlen\propto\xi$ minimizes
the cost per independent configuration and give a dynamical critical exponent
$z\approx1$. The coefficient of proportionality depends upon which operator is
being optimized, but it will typically be of order unity.

\subsection{Second Order Langevin Monte Carlo (L2MC)}

This algorithm was introduced by Horowitz \cite{horowitz90a} (and is sometimes
also called Kramers algorithm). It is similar to HMC except that only single
step (Langevin) trajectories are taken, and that instead of choosing the
fictitious momenta from a Gaussian heatbath after each step the final momenta of
a trajectory are reversed and mixed with a small amount of Gaussian noise to
serve as the initial momenta for the next trajectory.

The advantages are that because the trajectories are short they are not subject
to any large exponential amplification of rounding errors (which will be
discussed in this paper), and that the the free field theory mean acceptance
rate per trajectory is $\erfc(c\sqrt{V\dt^6})$ whereas for HMC it is
$\erfc(c'\sqrt{V\dt^4})$, where $V$ is the number of lattice sites, $\dt$ is the
leapfrog integration step size, and $c$ and $c'$ are constants of order unity
\cite{gausterer89a,gupta90a,kennedy91b,kennedy91a}.

The disadvantages are that an accurate CG solution is required for every
integration step (because each such step has an accept/reject test following
it), and that upon a rejection the trajectory reverses itself (because the
momenta are flipped). This last observation means that in order to minimize
autocorrelations the acceptance rate has to be very close to one with a
concomitantly tiny $\dt$.

A free field theory analysis \cite{kennedy91b,kennedy91a} indicates that L2MC
and HMC have essentially the same scaling behavior in both the thermodynamic
and continuum limits, so the choice between them must be made on the grounds of
the cost of specific implementations and on the possible costs associated with
instabilities occurring for long trajectories in HMC.

\subsection{Generalized Hybrid Monte Carlo (GHMC)}

The Generalized Hybrid Monte Carlo (GHMC) algorithm \cite{kennedy95a,kennedy91a}
combines the HMC and L2MC methods giving an algorithm with three tunable
parameters: the trajectory length $\trjlen$, the mixing angle between the old
fictitious momenta and Gaussian noise $\theta$, and the integration step size
$\dt$. A free field theory analysis indicates that if $\dt$ is chosen to given a
good acceptance rate then there is a valley in the $(\trjlen,\theta)$ plane for
which the dynamical critical exponent $z=1$ and furthermore for which the cost
per independent configuration does not vary much.

As HMC and L2MC are both special cases of GHMC, the optimal choice of parameters
may well not correspond to either case but in fact be somewhere in the middle of
this valley.

\section{Causes of Irreversibility}

Empirical studies \cite{edwards92a,jansen95a} have shown that HMC computations
for gauge theories are not exactly reversible. Given that they are performed
using floating point arithmetic this is not at all surprising. However, if $\dt$
or, for dynamical fermion calculations, the CG residual are taken to be too
large then very large violations of reversibility are observed. This large
violation of reversibility was observed even when the initial CG vector was
chosen to be zero.

We recall that the HMC algorithm is {\em exactly} reversible unless the initial
CG vector is chosen in a time asymmetric way\footnote{If the CG solution was
found exactly then it would of course be independent of the initial vector. In
practice only enough CG iterations are carried out to reduce the residual to
below some preset value, and the Krylov space thus explored depends on the
initial vector. If this initial vector depends on the solution for the previous
step then these Krylov spaces will be different for forward and backward
trajectories, and we will induce reversibility errors depending on the CG
residual used.} except for the effects of finite-precision arithmetic. We are
thus presented with the question of why the rounding errors are amplified by a
large factor.

Three possible causes of this amplification suggest themselves:
\begin{itemize}
  \item A small error changes the Metropolis accept/reject choice;
  \item A small error changes the number of CG iterations;
  \item The MD trajectories exhibit an exponential sensitivity to initial
  conditions (in other words they are unstable or
  chaotic) \cite{jansen95a,jansen96a}.
\end{itemize}
It is easy to see that although the first two mechanisms produce a large change
they occur correspondingly infrequently, and therefore are not compatible with
the nature of the irreversibility observed. We therefore conclude that the
irreversibility observed is caused by instabilities in the MD trajectories.

One may wonder to what extend irreversibility introduces systematic errors into
the results of an HMC calculation. If the irreversibility is due to some
underlying chaos in the equations of motion perhaps this just has the same
effect as a different choice of random numbers for the fictitious momentum
heatbath? Indeed, it is hard to see significant effects on physical observables
unless truly huge violations of reversibility are induced (in which case the
trajectories often become so unstable as to cause numerical overflows). We did
observe\footnote{See for example Figure~\ref{pe-ke-quenched} on
page~\pageref{pe-ke-quenched}.} that when the trajectories become irreversible
there is often a large change in the fictitious kinetic energy as well (even
though the total fictitious energy $\dH$ stays small). This means that the
distribution of fictitious momenta is clearly wrong, and thus the computation
is erroneous even if the effect on physically interesting observables is hard to
detect.

\section{Causes of Instability}

There are two distinct mechanisms which cause the MD equations to become
unstable:
\begin{itemize}
  \item The discrete integration procedure (leapfrog) diverges exponentially
  from solutions of the classical equations of motion. This instability should
  grow with the number of integration steps, and thus have a short
  characteristic time scale for large volumes.

  \item The classical equations of motion are themselves chaotic\footnote{Chaos
  in classical Yang-Mills dynamics (but not in fictitious time) has been
  investigated by several authors. See \cite{nielsen96a} and references cited
  therein.}. These may be expected to have a time scale independent of the step
  size, and presumably determined by some characteristic length scale of the
  system.
\end{itemize}

In section \ref{free-field-analysis} we analyze the instabilities of the
leapfrog integration scheme for free field theory, and in section \ref{chaos} we
consider the chaotic nature of the underlying continuous time equations of
motion themselves and propose a hypothesis as to how the time scale of the
corresponding instabilities behaves as the system approaches the continuum
limit.

\section{Free Field Theory Analysis}
\label{free-field-analysis}

The phenomenon of the leapfrog integration scheme diverging from the true
trajectory occurs even for free field theory, and it is instructive to see what
happens there \cite{kennedy91b,kennedy91a}.

Consider a system of harmonic oscillators $\{\phi_p\}$ for $p\in\Z_V$. The
Hamiltonian on fictitious phase space is
\begin{displaymath}
  H = \half\sum_{p\in\Z_V} \left(\pi_p^2 + \omega_p^2\phi_p^2\right).
\end{displaymath}

\subsection{Single Mode Analysis}

The Hamiltonian is diagonal, so we may temporarily consider the evolution of a
single mode and set its frequency $\omega_p=1$. In fact, we will find that the
essential features of the instability are displayed by a single mode.

The leapfrog discretization of Hamilton's equations can be written as the
following matrix acting on the phase space vector $(\phi, \pi)$
\begin{displaymath}
  U(\dt) = \twomatrix[1-\half\dt^2,\dt;-\dt+\quarter\dt^3,1-\half\dt^2].
\end{displaymath}
The general area-preserving reversible linear map on $(\phi,\pi)$ phase
space\footnote{If $F\equiv\twomatrix[1,0;0,-1]$ is the momentum-flip operation
then $F^{-1}U(\dt)F = U(-\dt)$, from which it immediately follows that $U_{1,1}$
and $U_{2,2}$ are even functions of $\dt$, and $U_{1,2}$ and $U_{2,1}$ are odd
functions of $\dt$. Furthermore $U(\dt)U(-\dt)=\identity$ implies that
$U_{1,1}=U_{2,2}$ except for the trivial case where $U_{1,2}=U_{2,1}=0$.} may be
parameterized as
\begin{displaymath}
  U(\dt) = \twomatrix[\cos[\kappa(\dt) \dt],
	  {\displaystyle \sin[\kappa(\dt) \dt] \over \displaystyle \rho(\dt)};
	-\rho(\dt) \sin[\kappa(\dt) \dt], {\cos[\kappa(\dt) \dt]}],
\end{displaymath}
where $\kappa$ and $\rho$ are even functions, but not necessarily real valued.
For the lowest-order leapfrog algorithm $\kappa$ and $\rho$ are
\begin{displaymath}
  \kappa(\dt) = {\cos^{-1}(1-\half\dt^2) \over\dt}, \qquad
  \rho(\dt) = \sqrt{1-\quarter\dt^2}.
\end{displaymath}

With this parameterization it is easy to see that the evolution operator for
a trajectory of $\trjlen/\dt$ leapfrog steps is
\begin{displaymath}
  U(\trjlen) = \twomatrix[\cos[\kappa(\dt) \trjlen],
	{\displaystyle \sin[\kappa(\dt) \trjlen] \over \displaystyle \rho(\dt)};
	-\rho(\dt) \sin[\kappa(\dt) \trjlen], {\cos[\kappa(\dt) \trjlen]}],
\end{displaymath}
and thus the change in fictitious energy over this trajectory is\footnote{The
subscript $p$ indicates that all fictitious times are to be multiplied by the
frequency~$\omega_p$.}
\begin{displaymath}
  \dH = {1\over2} \sum_{p\in\Z_V}
    \twovector[\phi_p, \pi_p]^T \Bigl[U_p^T U_p - \identity\Bigr]
      \twovector[\phi_p, \pi_p].
\end{displaymath}
It is clear that some kind of critical behavior will occur at $\dt=2$.

\begin{figure}
  \epsfxsize=0.7\textwidth
  \centerline{\epsfbox[0 50 288 238]{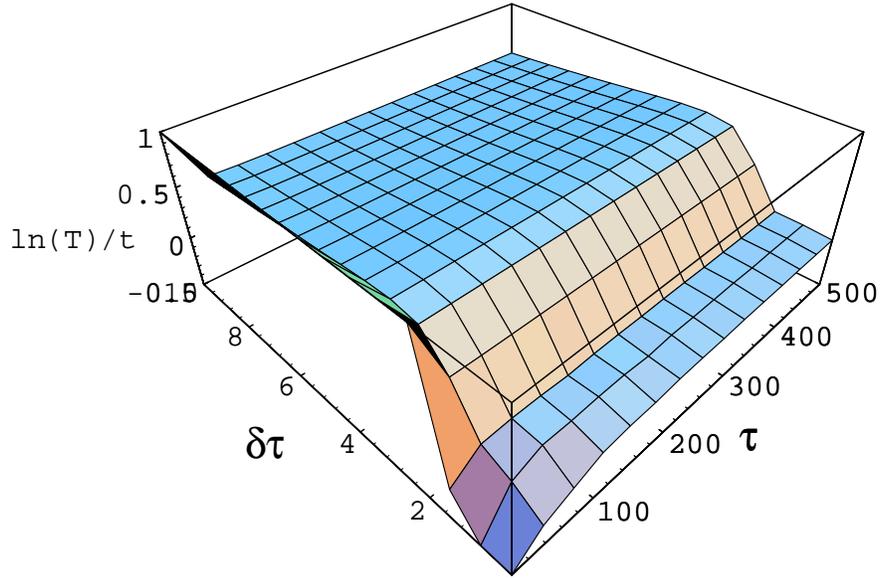}}
  \caption[Tplot]{The ``exponent'' $\Re\ln(T)/\trjlen$ is shown as a function of
    the integration step size $\dt$ and the trajectory length $\trjlen$. This
    quantity approaches $\nu$ as~$\trjlen\to\infty$.}
  \label{Tplot}
\end{figure}

Given this evolution matrix we can compute the probability distribution
of~$\dH$,
\begin{eqnarray}
  P_\dH(\xi)
    &=& {1\over Z} \int[d\phi][d\pi] e^{-H} \delta(\xi-\dH)
    = {1\over Z} \int[d\phi][d\pi] e^{-H}
      \int_{-\infty}^\infty {d\eta\over2\pi}\,e^{i\eta(\xi-\dH)} \nonumber \\
    &=& \int_{-\infty}^\infty {d\eta\over2\pi}\,{e^{i\eta\xi} \over
	\prod_p \sqrt{\det[\identity + i\eta(U_p^TU_p - \identity)]}}
    = \int_{-\infty}^\infty {d\eta\over2\pi}\,{e^{i\eta\xi} \over
	\prod_p \sqrt{1 + T_p^2i\eta(1-i\eta)}}
      \label{PdHIntegral-1}
\end{eqnarray}
where $T\equiv\left|\left(\rho-{\displaystyle1\over\displaystyle\rho}\right)
\sin(\kappa\trjlen)\right|$. It is easy to see that $T$ rapidly approaches its
asymptotic behavior of growing proportionally to $e^{\nu\trjlen}$ where
\begin{equation}
  \nu = \pm {1\over\dt}
    \Re\ln\left[\half\dt^2 - 1 \pm \dt\sqrt{\quarter\dt^2 - 1}\right],
  \label{single-mode-exponent}
\end{equation}
(which falls as $2\ln(\dt)/\dt$ as $\dt\to\infty$). For $\dt\leq2$ not only is
the exponent $\nu$ zero, but $T$ is bounded by a constant. That the exponent
decreases as $\dt\to\infty$ merely reflects the fact that $T$ is growing
exponentially in the number of MD steps but only algebraically in the step
size~$\dt$. This is illustrated in Figure~\ref{Tplot}.

\begin{figure}
  \epsfxsize=0.7\textwidth
  \centerline{\epsfbox[0 50 288 238]{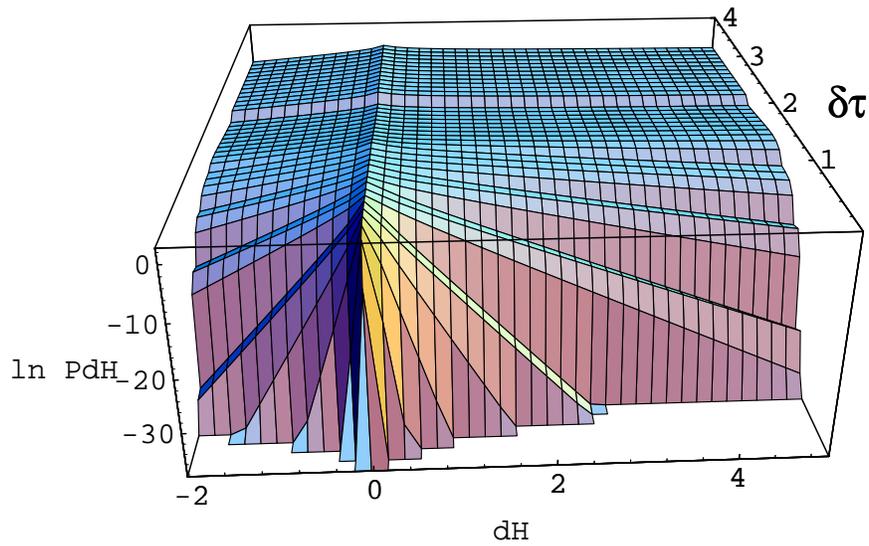}}
  \caption[PdHplot]{The logarithm of the probability distribution for the change
    in fictitious energy $\dH$ for a single mode is shown as a function of $\dH$
    and the MD leapfrog integration step size $\dt$ for a trajectory of length
    $\trjlen=4$ (rounded to the nearest integer multiple of~$\dt$).}
  \label{PdHplot}
\end{figure}

For a single mode the probability distribution of $\dH$ may be evaluated by
changing variable to $\eta'\equiv\eta+i/2$
\begin{equation}
  P_\dH(\xi)
    = \int_{-\infty}^\infty {d\eta\over2\pi}\,
	{e^{i\eta\xi} \over \sqrt{1 + T^2i\eta(1-i\eta)}}
    = e^{\xi/2} \int_{i/2}^{\infty+i/2} {d\eta'\over\pi}\,
	{\cos(\eta'\xi) \over \sqrt{1 + T^2(\quarter+\eta'^2)}}.
      \label{PdHIntegral-2}
\end{equation}
The singularities in the integrand occur when $1 + T^2(\quarter+\eta'^2)=0$, for
which $\eta'^2<-\quarter$ and hence $|\Im\eta'|>\half$. We may thus shift the
integration contour to the real axis, and defining $x\equiv T\eta'/\sqrt{1 +
\quarter T^2}$ we obtain
\begin{displaymath}
  P_\dH(\xi) = {e^{\xi/2}\over\pi T} \int_0^\infty {dx\over\sqrt{1+x^2}}
    \cos\left({x\xi\over2}\sqrt{1+{4\over T^2}}\right),
\end{displaymath}
which can be expressed in closed form in terms of a modified Bessel function
\begin{displaymath}
  P_\dH(\xi) =
    {e^{\xi/2}\over\pi T} K_0\left({|\xi|\over2}\sqrt{1 + {4\over T^2}}\right).
\end{displaymath}
The logarithm of the acceptance rate is shown in Figure~\ref{PdHplot}.

\begin{figure}
  \epsfxsize=0.7\textwidth
  \centerline{\epsfbox[0 50 288 238]{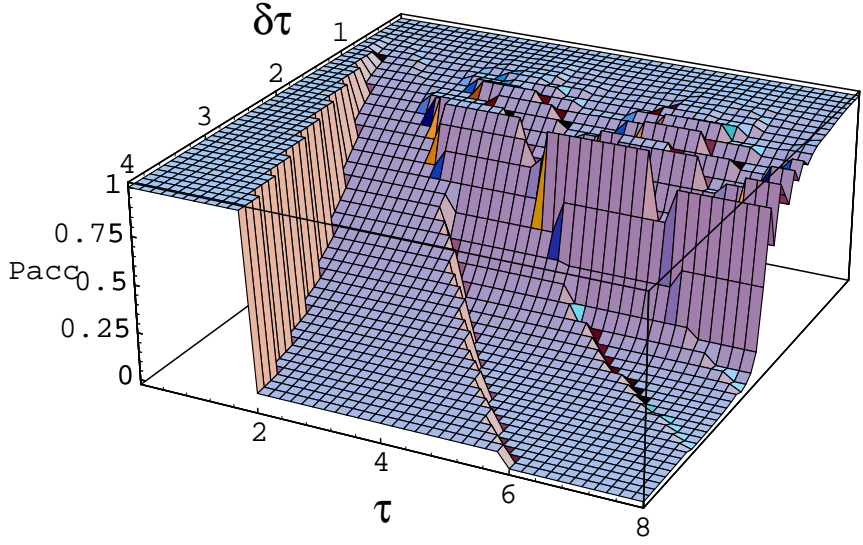}}
  \caption[accplot]{Acceptance rate for a single mode as a function of the MD
    step size $\dt$ and trajectory length $\trjlen$ (rounded to the nearest
    integer multiple of $\dt$).}
  \label{accplot}
\end{figure}

The average Metropolis acceptance rate is
\begin{equation}
  \pacc \equiv \Big\langle\min(1, e^{-\dH})\Big\rangle
    = \int_{-\infty}^\infty d\xi\,\min(1,e^{-\xi})P_\dH(\xi).
  \label{Pacc-def}
\end{equation}
For a single mode we may use eq.~(\ref{PdHIntegral-2}), and evaluating the
integral over~$\xi$ gives
\begin{displaymath}
  \pacc = {2\over\pi T} \int_{-\infty}^\infty {dx\over\sqrt{1+x^2}
    \left[1+x^2\left(1+{4\over T^2}\right)\right]};
\end{displaymath}
setting $x\equiv{1\over2}\left(\sqrt{y} - {\displaystyle1\over \displaystyle
\sqrt{y}}\right)$ and ${2\over T} \equiv \tan{\theta\over2}$ we easily obtain
\begin{displaymath}
  \pacc = {\sin\theta\over\pi} \int_0^\infty {dy\over y^2 + 2y\cos\theta + 1}
%%    = {i\over2\pi} \int_0^\infty dy
%%      \left\{ {1\over y+e^{i\theta}} - {1\over y+e^{-i\theta}} \right\}
    = {\theta\over\pi} = {2\over\pi} \tan^{-1}\left({2\over T}\right),
\end{displaymath}
and this is shown in Figure~\ref{accplot} as a function of the step size and
trajectory length. Observe that there is a ``wall'' at $\dt=2$ which separates
the region where the acceptance rate oscillates as a function of $\trjlen$ from
the region where it plummets exponentially.

\subsection{Multiple Mode Analysis}

These results can be generalized to a field theory with many ``stable'' modes
and some number of ``unstable'' ones. In the general case let us consider the
acceptance rate $\pacc$ defined in eq.~(\ref{Pacc-def}) and use the expression
of eq.~(\ref{PdHIntegral-1}) for the distribution of values of~$\dH$. As before
we may make the change of variable $\eta'\equiv\eta+i/2$ and observe that there
are no singularities in the strip of the complex~$\eta'$ plane satisfying
$|\Im\eta'|<\half$. We thus obtain
\begin{eqnarray*}
  \pacc &=& \int_{-\infty}^0 d\xi\,P_\dH(\xi) +
      \int_0^\infty d\xi\,P_\dH(\xi) e^{-\xi} \\
    &=& \int_{-\infty}^\infty {d\eta'\over2\pi}\,
      {1\over\prod_p\sqrt{1 + T_p^2(\quarter+\eta'^2)}}
      \left[\int_{-\infty}^0 d\xi\,e^{(i\eta'+\half)\xi} +
	\int_0^\infty d\xi\,e^{(i\eta'-\half)\xi}\right] \\
    &=& \int_{-\infty}^\infty {d\eta'\over2\pi}\,
      {1\over(\quarter+\eta'^2)\prod_p\sqrt{1 + T_p^2(\quarter+\eta'^2)}}.
\end{eqnarray*}
Let us now assume that there are a small number $N$ of ``unstable'' modes with
$T_q\gg1$ and a large number $V$ of ``stable'' modes with $T_p\ll1$ such
that\footnote{If $\dt\ll1$ then $T_p = -\quarter(\omega_p\dt)^2
\sin(\omega_p\trjlen) + O(\dt^4)$, and thus $\sum_{p=1}^V T_p^2 = kV\dt^4 +
O(\dt^6)$ where the spectral average $k\equiv {1\over16V} \sum_{p=1}^V
\omega_p^4 \sin^2(\omega_p\trjlen)$ is of order unity. In order to obtain a
non-negligible acceptance rate we must take $\dt$ to be of order
$V^{-\quarter}$, in which case the quantity $\sigma$ will also be of order one.}
$\sum_{p=1}^V T_p^2\equiv \sigma^{-2}$ and $\sum_{p=1}^V T_p^{2n} = O(V^{1-n})$.
We can thus compute an asymptotic expansion valid for large $V$,
\begin{eqnarray*}
  \pacc &=& \int_{-\infty}^\infty {d\eta'\over2\pi}\,
      {\exp\left\{-\half\sum_{p=1}^V
	  \ln\left[1 + T_p^2(\quarter+\eta'^2)\right]\right\}
	\over(\quarter+\eta'^2)
	  \prod_{q=1}^N \sqrt{1 + T_q^2(\quarter+\eta'^2)}} \\
    &\sim& e^{-1/8\sigma^2} \int_{-\infty}^\infty {d\eta'\over2\pi}\,
      {e^{-\eta'^2/2\sigma^2}
	\over(\quarter+\eta'^2)\prod_q\sqrt{1 + T_q^2(\quarter+\eta'^2)}}
      \biggl(1 + \quarter\sum_p T_p^4 (\quarter+\eta'^2)^2 + \cdots\biggr),
\end{eqnarray*}
where the ``stable'' modes correspond to subscript $p$ and the unstable ones to
subscript~$q$. If there are a few modes with $V^{-1/4}\ll\omega_p\dt\ll1$ then
these may be taken into account by computing further terms in the asymptotic
expansion above. On the other hand, we may carry out an expansion in powers of
$1/T_q$ for the ``unstable'' modes by writing
\begin{equation}
  {1\over\prod_{q=1}^N\sqrt{1 + T_q^2(\quarter+\eta'^2)}} =
    {(\quarter+\eta'^2)^{-N/2}\over\prod_q T_q}
      \biggl(1 - \half\sum_q {1\over T_q^2} (\quarter+\eta'^2)^{-1} +
	\cdots\biggr).
\end{equation}
Using this expansion in the integrand above reduces both the large $T$ and small
$T$ expansions to a sum of integrals of the form
\begin{equation}
  \int_0^\infty d\eta'\, e^{-\eta'^2/2\sigma^2} (\quarter+\eta'^2)^{-k/2}
    = 2^{k-2} \sqrt\pi U\Bigl({1\over2}, {3-k\over2}; {1\over8\sigma^2}\Bigr)
\end{equation}
where $U$ is a confluent hypergeometric function.\footnote{This is related to
Whittaker's function $W_{\lambda,\mu}(z) = z^{\mu+\half} e^{-z/2}
U(\mu-\lambda+\half, 2\mu+1; z)$.} It is easily verified that for $N=0$ the
leading term in the asymptotic expansion is just $\erfc(1/\sqrt{8\sigma^2})$.

\begin{figure}
  \epsfxsize=0.7\textwidth
  \centerline{\epsfbox[0 50 288 238]{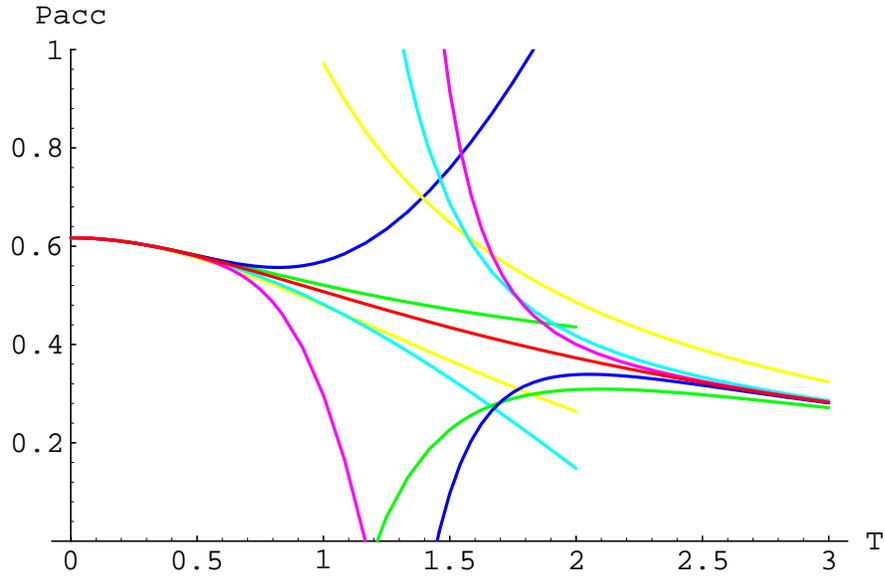}}
  \caption[t-mode-plot]{Acceptance rate for many ``stable'' modes with
    $\sigma=1$ and one ``unstable'' mode as a function of the value of $T$ for
    the ``unstable'' mode. In addition to the result obtained by numerical
    integration various orders of the small- and large-$T$ expansions are
    shown.}
  \label{t-mode-plot}
\end{figure}

\begin{figure}
  \epsfxsize=0.7\textwidth
  \centerline{\epsfbox[0 50 288 238]{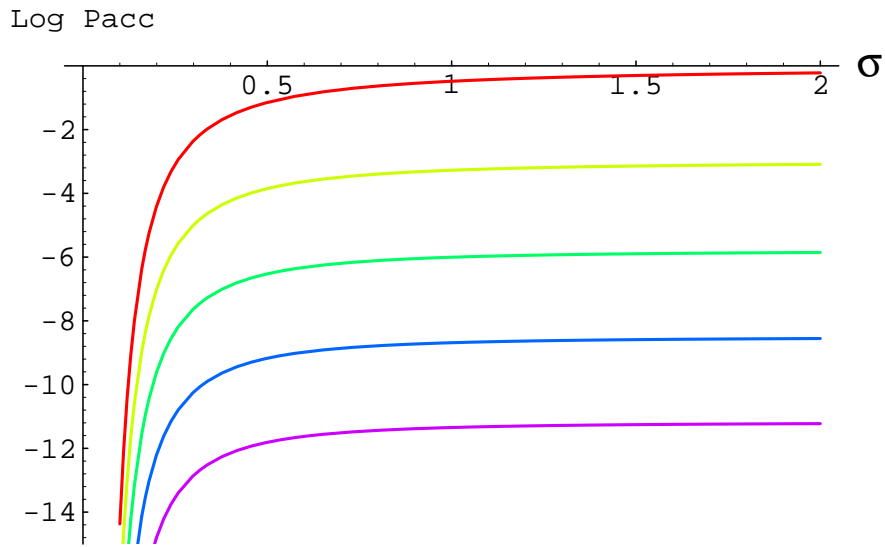}}
  \caption[pacc-n-2d]{Logarithm of the mean acceptance rate for $N$ ``unstable''
    modes as a function of $\sigma$ for $T=25.63$ which corresponds to $\dt=2.1$
    and $\trjlen=9$. The four curves are for $N$ from $0$ (top) to $4$
    (bottom).} \label{pacc-n-2d}
\end{figure}

\begin{figure}
  \epsfxsize=0.7\textwidth
  \centerline{\epsfbox[0 50 288 238]{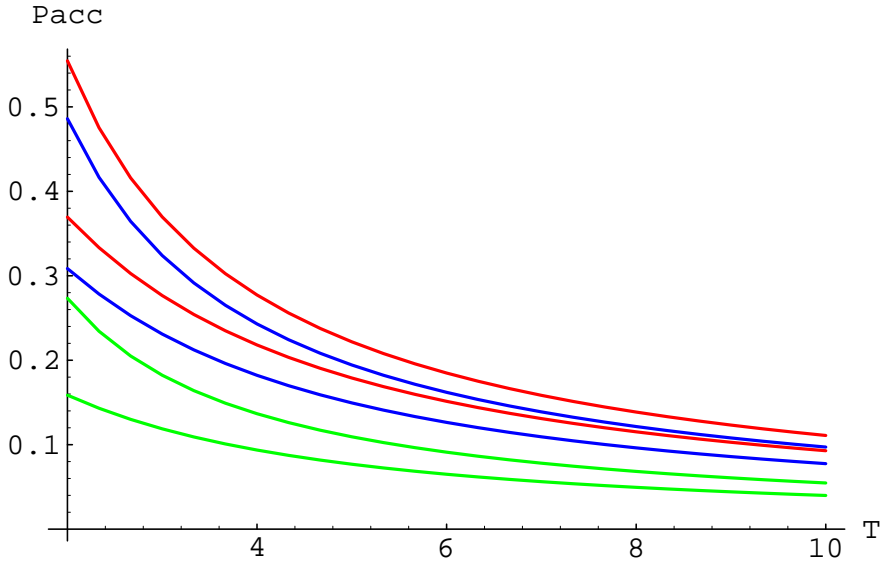}}
  \caption[single-multiple]{Comparison of the acceptance rate for a single mode
    (multiplied by a factor of $\erfc(1/\sqrt{8\sigma^2})$ compared with the
    corresponding acceptance rate for one ``unstable'' mode and many ``stable''
    modes for $\sigma=0.5$, $1$, and~$1.5$.}
  \label{single-multiple}
\end{figure}

In Figure~\ref{t-mode-plot} we show the $\pacc$ for the case where there are
many ``stable'' modes with $\sigma=1$ as a function of the value of $\sigma$ for
one additional ``unstable'' mode. It is clear that the expansions calculated
above are very good as long as $T$ is not close to one.
In Figure~\ref{pacc-n-2d} we plot $\ln\pacc$ as a function of $T$ for various
values of $N$, the number of ``unstable'' modes. Each additional unstable mode
clearly reduces the acceptance rate by a large factor for these parameters, and
this makes it reasonable to assume that at least the qualitative behavior of
the instabilities in the system is apparent from the case where there is just a
single unstable mode.
In Figure~\ref{single-multiple} we show a comparison of the dependence of the
mean acceptance rate on $T$ for a single mode and for a system of many
``stable'' modes and one ``unstable'' one. This demonstrates that the behavior
of a single mode dominates the instabilities in the case of free field theory.

We notice that in order to keep the acceptance rate constant as the lattice
volume $V\to\infty$ we must decrease $\dt$ so as to keep $V\dt^4$ fixed. Thus as
we approach the thermodynamic limit the instabilities will go away: in other
words the leapfrog instability is a finite volume effect.

\subsection{Interacting Field Theories}

When interactions are present it is no longer meaningful to talk of independent
modes, but for weak interactions like those for asymptotically free field
theories at short distance it is still useful to consider the system as a set of
weakly coupled modes. We expect that the onset of the integration instability
will still be caused by the highest frequency mode, and will occur
at~$\omega_{\max}\dt=2$. The forces acting on the highest frequency mode due to
the other modes will fluctuate in some complicated way, and consequently we
would expect that the ``wall'' at $\omega_{\max}\dt=2$ gets smeared out.

\begin{figure}
  \epsfxsize=0.7\textwidth
  \centerline{\epsfbox{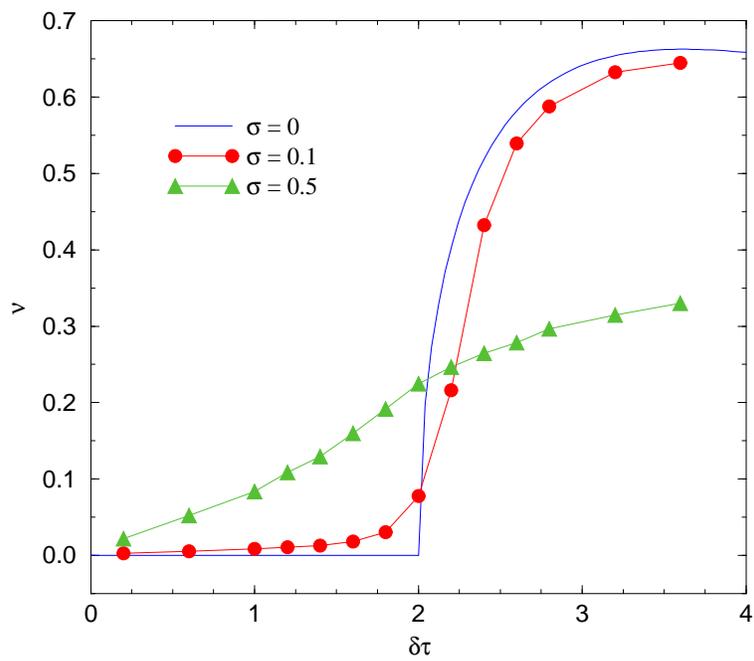}}
  \caption[fluc]{Results from fluctuating-frequency harmonic oscillator model.
    The curve for $\sigma=0$ is given by eq.~(\ref{single-mode-exponent}) on
    page~\pageref{single-mode-exponent}.}
  \label{fluc}
\end{figure}

In order to illustrate this effect  we have made numerical studies of a simple
model of a single harmonic oscillator whose frequency is randomly chosen from a
Gaussian distribution with mean unity and standard deviation~$\sigma$ before
each MD step. The behavior of this model is just like that of the free field
theory considered above, except that the ``wall'' at $\dt=2$ gets spread out.
Our numerical results are shown in Figure~\ref{fluc}, where the analytic result
for $\sigma=0$ is taken from eq.~(\ref{single-mode-exponent}).
We conjecture that interacting asymptotically free field theories like
non-Abelian gauge theories behave similarly, and this is supported by our
numerical results.

We do not expect the introduction of dynamical fermions to produce any
qualitatively different behavior. For QCD computations using the HMC algorithm
the pseudofermions produce a force which we expect to be proportional to
$1/m_f$, so the ``highest frequency'' of the system, which is responsible for
the integration instability, will grow as $\omega_{\max}\propto1/m_f$ too. In
particular we expect $\omega_{\max}$ to become large as $m_f\to0$, so the
critical value of $\dt$ will become small as $\kappa\to\kappa_c$ for Wilson
fermions. An inaccurate CG solution will produce errors which are multiplied by
the scale of this fermionic force, and will become important when the size of
the errors in the CG solution become comparable with the size of the
fluctuations in the forces due to the fields themselves.

It is interesting to observe that reversibility can be violated significantly
even though the acceptance rate does not become small. For example we will see
an example of this in Figure~\ref{tau-ddu-dh} on page~\pageref{tau-ddu-dh} for
$\dt=0.005$ and~$\trjlen=50$. This is because the leapfrog integration scheme
tries to conserve energy even when it has wandered far away from the true
continuous time trajectory. It is therefore prudent to verify that reversibility
is satisfied to the precision required, even when the HMC algorithm has an
average acceptance rate which is close to one.

\section{Chaos in Continuous Time Evolution}
\label{chaos}

Jansen and Liu \cite{jansen95a} observed that there is a second cause of
instability in the MD trajectories, which occurs even when the integration step
size~$\dt\to0$. This is because the underlying continuous fictitious time
equations of motion are themselves chaotic. In the chaotic regime two nearby
classical trajectories diverge exponentially with a characteristic exponent
$\nu$ called the Liapunov exponent. Obviously such behavior cannot be studied
in the context of free field theory.

The instability due to chaos cannot be removed by reducing $\dt$, and is
therefore not a finite size effect. If the system we are studying exhibits
critical slowing down,\footnote{This is probably not the case for most current
lattice QCD computations.} and we believe that the mechanism for reducing this
to $z=1$ in free field theory is applicable to this system too, and we wish to
use the HMC algorithm, then we must scale the trajectory length with the
correlation length, $\trjlen\propto\xi$. This means that there is the potential
for exponential amplification of rounding errors.

Even if there is such an exponential amplification of rounding errors,
reversibility can be maintained to any desired precision by a linear increase in
the number of digits used for floating point arithmetic. This is no longer a
negligible addition to the computational complexity of the problem (i.e., it is
not just a logarithmic correction), but it is still a small correction to the
power dependence of the computational cost on the correlation length and volume.

This problem can be avoided using the GHMC algorithm, but not only is this
unnecessary until chaotic amplification of rounding errors becomes important,
but also as we shall see it might never be necessary at all.

The reason for this is that the Liapunov exponent (averaged over the equilibrium
distribution) is not constant as a function of $\beta$. If the chaotic dynamics
is not only a property of the underlying continuous fictitious time evolution,
but is also a property of the underlying (space-time) continuum field theory,
then the Liapunov exponent would be constant when measured in ``physical''
units, that is $\nu\xi$ would be constant as $\xi\to\infty$. In other words this
hypothesis would say that $\nu \propto 1/\xi$, or for QCD with $n_f$ flavors of
fermions
\begin{equation}
  \nu \propto e^{-\beta/12\beta_0}
    \qquad \hbox{\rm with} \qquad \beta_0={11-{2\over3}n_f\over16\pi^2},
  \label{asymptotic-scaling}
\end{equation}
where the last relation follows from perturbation theory for $\beta$ large
enough that the system exhibits asymptotic scaling. In
section~\ref{liapunov-parameter-dependence} we shall present numerical evidence
that this hypothesis may indeed be true. If this is the case, then tuning the
HMC algorithm by varying the trajectory length proportionally to the correlation
length does not lead to any change in the amplification of rounding errors as we
change~$\xi$.

\section{Monte Carlo Results}

We have carried out extensive numerical studies of reversibility errors for
(quenched) $SU(3)$ gauge theory and for full QCD with two flavors of dynamical
Wilson fermions. Measurements were made principally on $4^4$ lattices, but
additional data on $8^4$ lattices was used for an analysis of finite size
effects.

\subsection{Numerical Determination of Reversibility Errors}

In order to measure the deviation from reversibility of MD trajectories we chose
some initial point in fictitious phase space $(U_i,\pi_i)$ and followed an MD
trajectory of length $\trjlen$ to the final point $(U_f,\pi_f)$; we then
reversed the fictitious momenta and followed the backward trajectory of length
$\trjlen$ from $(U_f,-\pi_f)$ to $(U'_i,\pi'_i)$. At the end of this backward
trajectory we measured the amount by which it failed to return to its starting
point by computing two quantities, the change in energy $\Dd H\equiv
H(U_i',\pi_i') - H(U_i,\pi_i)$, and the norm of the change in the gauge
field\footnote{The configuration norm $||\Dd U||$ could have been made into a
phase space norm by adding $||\Dd \pi||$ to it, but we do not believe that this
would make any significant difference in practice (and it would  require saving
the initial momenta $\pi_i$, which are not needed in the usual HMC algorithm but
are needed for the L2MC and GHMC algorithms).} $||\Dd U||\equiv ||U_i' - U_i||$,
where $||U||^2 \equiv \sum_{x,\mu} \sum_{a,b} \Bigl|[U_\mu(x)]_{ab}\Bigr|^2$,
$a$ and $b$ being $SU(3)$ color indices. Observe that both quantities are
extensive in the lattice volume, that $\Dd H$ is gauge invariant whereas $||\Dd
U||$ is not, and that $||\Dd U||$ cannot be small unless $\Dd H$ is too. These
quantities can also be considered as the difference between the change in energy
or gauge field over the forward and the backward trajectories, which serves to
explain our notation. Both of these quantities vanish identically for reversible
trajectories, of course.

\subsubsection{$SU(3)$ Gauge Theory}

\begin{figure}
  \epsfxsize=0.7\textwidth
  \centerline{\epsfbox{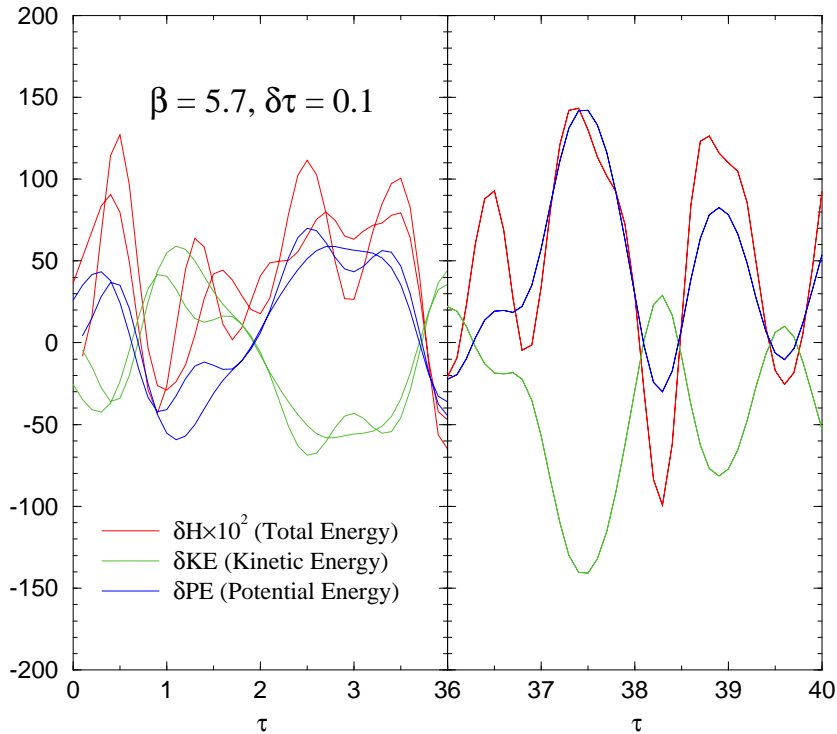}}
  \caption[pe-ke-quenched]{The contributions to $\dH$ from the fictitious
    kinetic energy and the potential energy (gauge action) for a trajectory of
    length $\trjlen=40$ with step size $\dt=0.1$ at $\beta=5.7$. The two graphs
    show the contributions to the energy for the first and last four units of
    fictitious time respectively, overlaid with the corresponding quantities on
    the backward trajectory. The graph on the right shows that the forward are
    backward trajectories are essentially indistinguishable immediately after
    the reversal, and the graph on the left shows how the backward trajectory
    finally deviates from the forward one. Note that the cancellation between
    $\delta$KE and $\delta$PE is still at the 2\% level even at the end of this
    very long trajectory with large step size, as the $\dH$ values have been
    multiplied by a factor of~100.}
  \label{pe-ke-quenched}
\end{figure}

For $SU(3)$ gauge theory an example of a long trajectory is shown in
Figure~\ref{pe-ke-quenched}. This shows how the energy $\dH$ changes along a
long trajectory. The change in energy over the forward trajectory was
$\dH=1.2$ (corresponding to an acceptance probability of 30\%), which differed
from the change in energy over the backward trajectory by $\Dd H=0.36$. This
clearly shows that the leapfrog integration scheme tries very hard to conserve
energy even though it has wandered far from the true path: indeed, the
backward trajectory ends up a distance $||\Dd U||=20$ away from the starting
configuration.
This example demonstrates that there can be severe violations of reversibility
even though the acceptance rate is quite reasonable, but it should be noted that
a trajectory length of $\trjlen=40$ is unreasonably long for a system whose
correlation length $\xi=O(1)$, especially on a $4^4$ lattice!

\begin{figure}
  \epsfxsize=0.7\textwidth
  \centerline{\epsfbox{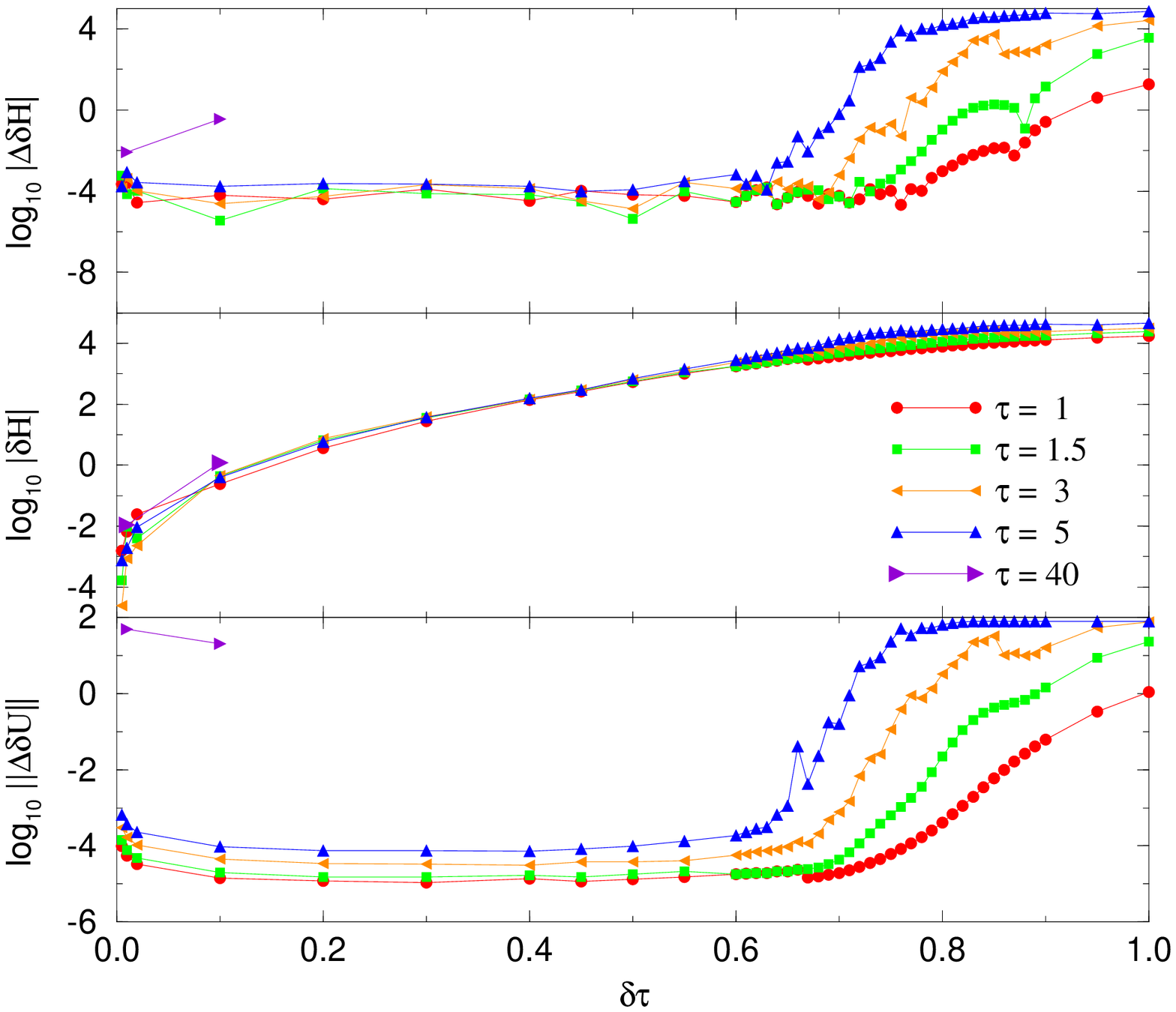}}
  \caption[dt-ddu-dh-ddh]{Results for pure $SU(3)$ gauge theory with $\beta=5.7$
    on a $4^4$ lattice as a function of $\dt$.}
  \label{dt-ddu-dh-ddh}
\end{figure}

\begin{figure}
  \epsfxsize=0.7\textwidth
  \centerline{\epsfbox{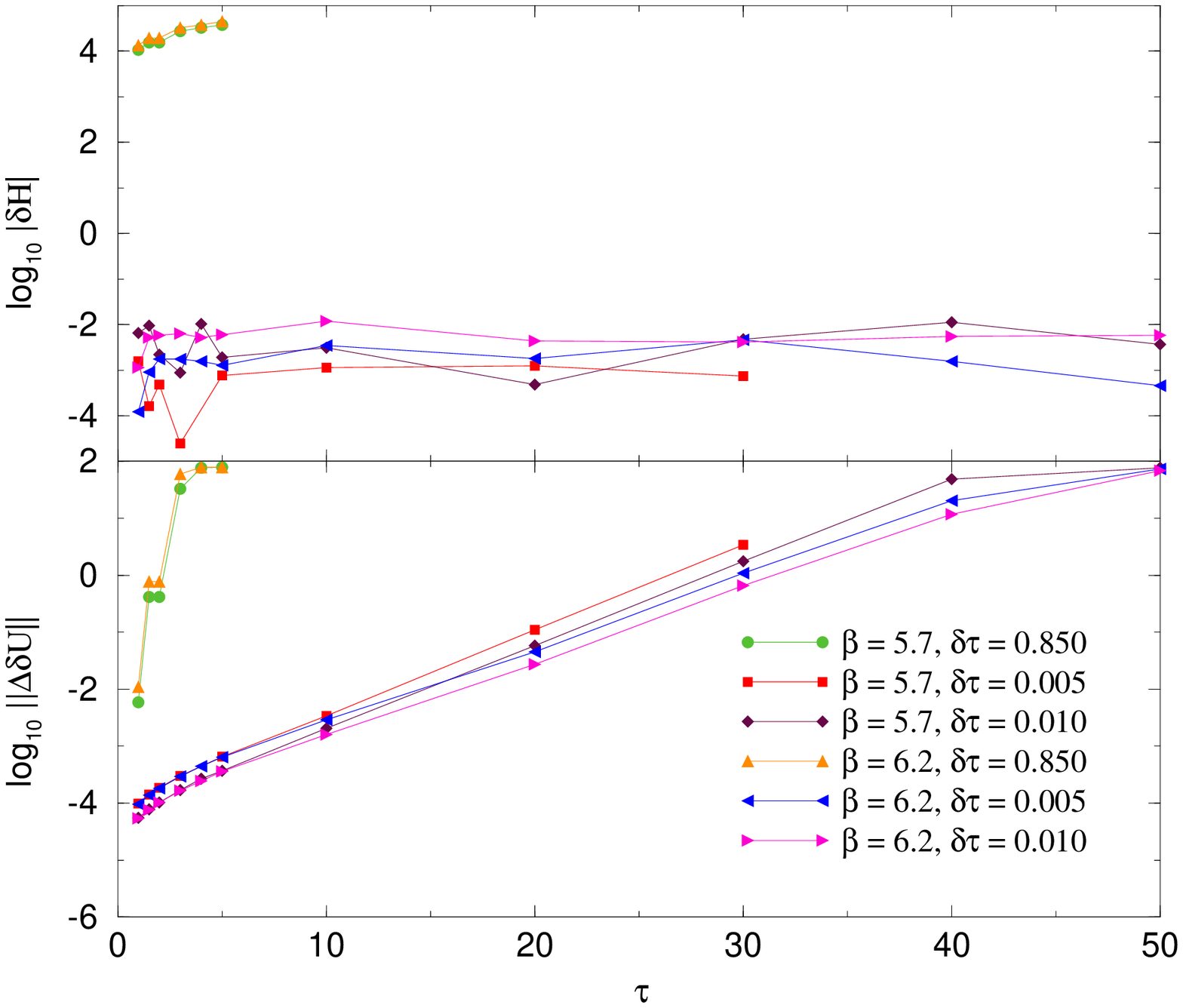}}
  \caption[tau-ddu-dh]{Results for pure $SU(3)$ gauge theory with $\beta=5.7$
    and $6.2$ on a $4^4$ lattice as a function of $\trjlen$.}
  \label{tau-ddu-dh}
\end{figure}

By carrying out such forward and backward trajectories for a variety of step
sizes and trajectory lengths we produced Figures~\ref{dt-ddu-dh-ddh}
and~\ref{tau-ddu-dh}.
In the former we plot the logarithm of $|\Dd H|$, $|\dH|$, and $||\Dd U||$ as a
function of the integration step size. The differences between measurements made
on different equilibrated configurations were very small. The ``kinks'' in some
of the curves are just a consequence of rounding $\trjlen$ to the nearest
integer multiple of $\dt$. The top and bottom graphs clearly show the
integration instability ``wall'' at $\dt\approx0.6$, which has spread out just
as we would expect. The middle graph shows that by the time one has reached the
``wall'' $\dH=O(10^3)$, so the integration instabilities are of no practical
importance for this system. If one looks carefully at the left end of the bottom
graph one sees that $||\Dd U||$ increases a little as $\dt\to0$. This is because
the number of integration steps $n=\trjlen/\dt$ is increasing, and therefore
rounding errors are being amplified by the usual $\sqrt n$ factor.

For extremely long trajectories reversibility is violated even for very small
values of $\dt$; in order to understand this it is instructive to look at
Figure~\ref{tau-ddu-dh}, where we plot the logarithm of $\dH$ and $||\Dd U||$ as
a function of the trajectory length for two values of $\beta$ and three values
of~$\dt$. The lower graph immediately shows us that even for very small step
sizes the errors are amplified exponentially, with an exponent which is
independent of $\dt$ (i.e., the lines for the same $\beta$ value but different
$\dt$ are parallel). This is evidence for chaos in the continuous time equations
of motion, as suggested by Jansen and Liu \cite{jansen95a}. The lines for small
step size on the lower graph eventually curve downwards for the largest
$\trjlen$ values because $SU(3)$ is a compact manifold so there is a maximum
distance two configurations can be from one another, and this bound is becoming
saturated. The upper graph shows that $\dH$ stays small for small $\dt$ even
though the rounding errors have been enormously amplified.

\subsubsection{QCD with Wilson Fermions}

We have carried out the corresponding analysis for full QCD with two flavors of
Wilson fermions with configurations taken from two ensembles, one at $\beta=5.1$
and $\kappa=0.16$ which corresponds to quite heavy quarks, and one at
$\beta=3.5$ and $\kappa=0.2225$ which is extremely close to $\kappa_c$. We took
care that both ensembles were in the ``confined phase,'' if we use thermodynamic
terminology to approximately describe the finite size effects on our symmetric
lattices.

\begin{figure}
  \epsfxsize=0.7\textwidth
  \centerline{\epsfbox{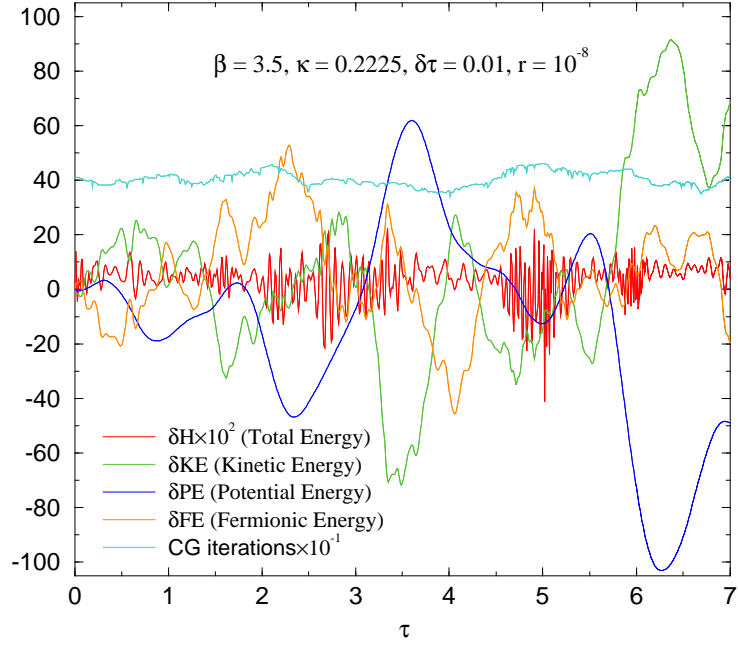}}
  \caption[pe-ke-dynamical]{The contributions to $\dH$ from the fictitious
    kinetic energy, potential energy (gauge action), and the fermionic energy
    for a trajectory of length $\trjlen=7$ with step size $\dt=0.01$ at
    $\beta=3.5$ and $\kappa=0.2225$. Each curve is overlaid with the
    corresponding quantities on the backward trajectory, but the forward and
    backward trajectories are indistinguishable on this scale. Note that $\dH$
    has been scaled by a factor of 100 because it has a magnitude of less than a
    percent of its constituent parts. The number of CG iterations used to reach
    the residual $r=10^{-8}$ from a zero start is also shown (divided by a
    factor of~10).}
  \label{pe-ke-dynamical}
\end{figure}

Figure~\ref{pe-ke-dynamical} shows a typical trajectory for the light dynamical
fermion system. For this trajectory $\dH=3.9\times10^{-2}$ corresponding to an
acceptance probability of 96\%, and $||\Dd U||=1.4\times10^{-2}$. This
trajectory exhibits some typical behavior: the potential energy oscillates
smoothly with a period of about $2$ MD time units; the fermionic energy is much
more jittery; and the kinetic energy adjusts itself to cancel the other two
components leaving a small residual $\dH$ whose magnitude is about a percent of
that of its constituent parts. The three constituent contributions to the energy
are of similar magnitudes, which indicates that the fermions are indeed very
light: with heavier fermions the fermionic energy is much smaller.

\begin{figure}
  \epsfxsize=0.7\textwidth
  \centerline{\epsfbox{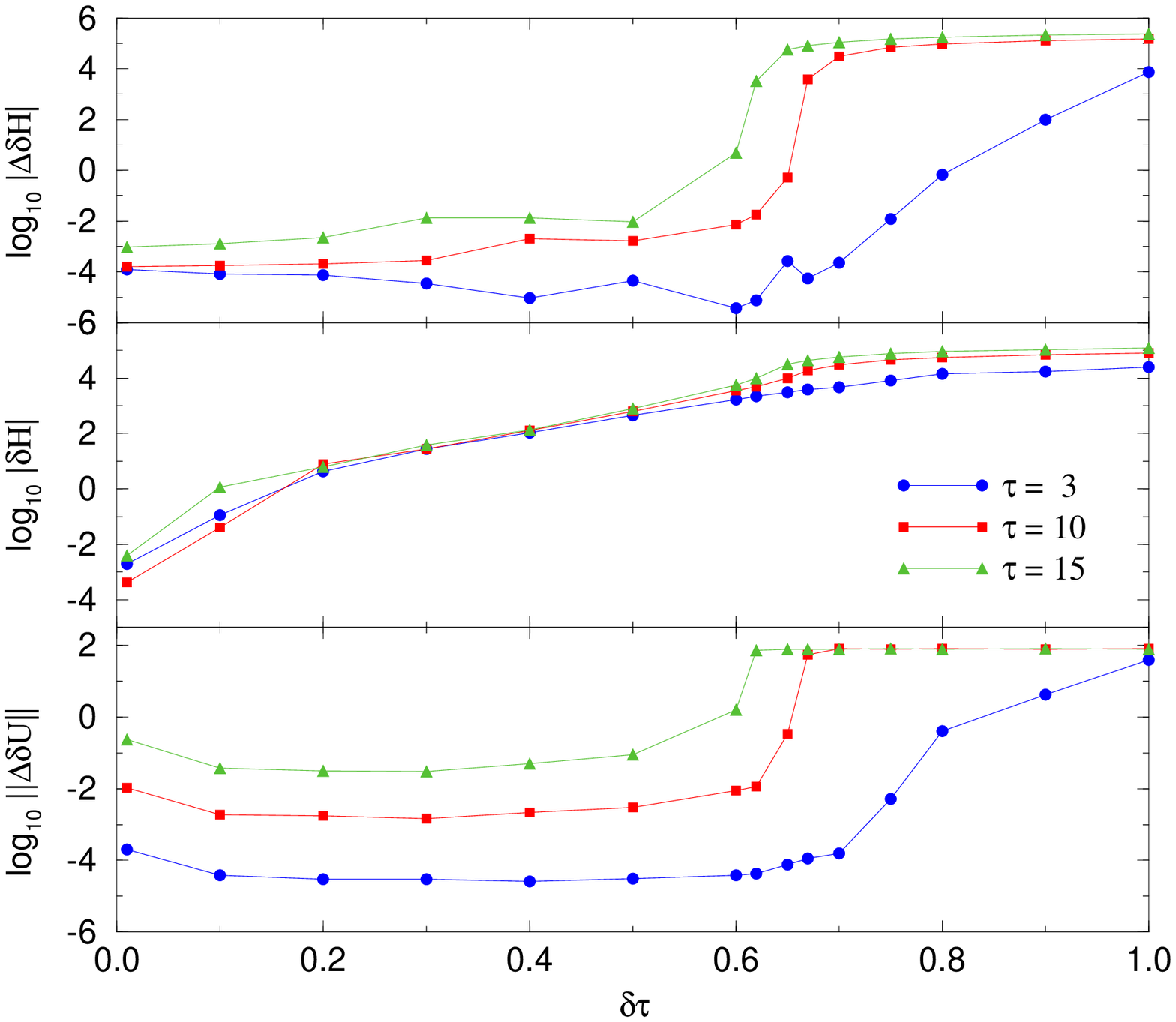}}
  \caption[dt-ddu-dh-ddh-dyn]{Results for QCD as a function of integration step
    size $\dt$. The data shown is for $\beta=5.1$, $\kappa=0.16$, and CG
    residual $r=10^{-8}$ on a $4^4$~lattice.}
  \label{dt-ddu-dh-ddh-dyn}
\end{figure}

\begin{figure}
  \epsfxsize=0.7\textwidth
  \centerline{\epsfbox{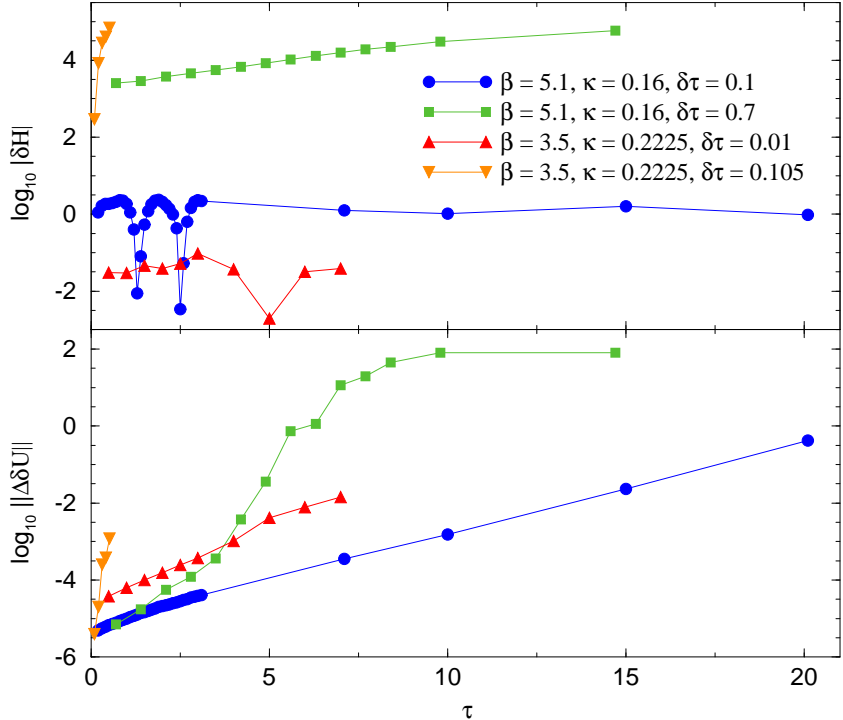}}
  \caption[tau-ddu-dh-dyn]{Results for QCD as a function of trajectory length
    $\trjlen$. The data shown is from $4^4$~lattices with a CG residual
    $r=10^{-8}$.}
  \label{tau-ddu-dh-dyn}
\end{figure}

\begin{figure}
  \epsfxsize=0.7\textwidth
  \centerline{\epsfbox{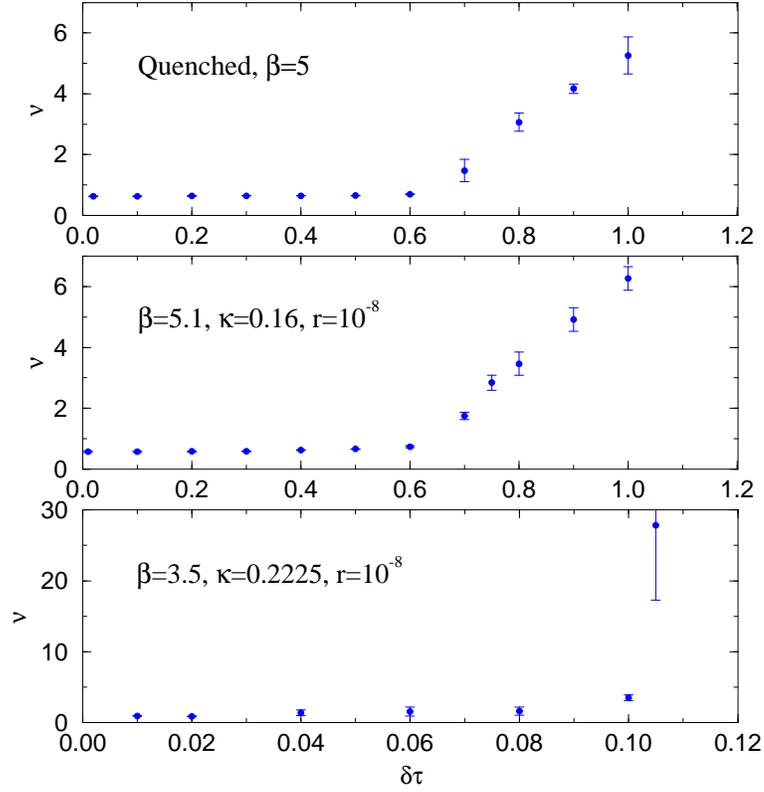}}
  \caption[exponent-dt]{The Liapunov exponent $\nu$ is shown as a function of
    the integration step size $\dt$. The top graph is for pure $SU(3)$ gauge
    theory, the middle one is for QCD with heavy dynamical Wilson quarks, and
    the bottom one is for QCD with light dynamical Wilson quarks. Note that the
    scale for the light quark case is quite different from the other two. The
    error bars show the standard deviation of measurements made on three
    independent configurations.}
  \label{exponent-dt}
\end{figure}

From many such trajectories on several equilibrated configurations we produced
the data shown in Figures~\ref{dt-ddu-dh-ddh-dyn} and~\ref{tau-ddu-dh-dyn}. For
simplicity\footnote{The data for the light fermions is similar in form but on a
very different $\dt$ scale. The results obtained from all the data will be shown
in Figure~\ref{exponent-dt} on page~\pageref{exponent-dt}.} we only show the
data for the heavy fermion case in Figure~\ref{dt-ddu-dh-ddh-dyn}. The results
are very similar to the pure $SU(3)$ gauge theory results of
Figures~\ref{dt-ddu-dh-ddh} and~\ref{tau-ddu-dh}, especially if one notes that
the trajectory lengths are somewhat longer in the present case.
Just as in that case we observe the smeared-out ``wall'' at which the leapfrog
integration scheme becomes unstable, but that the acceptance rate is already
completely negligible by the time the wall is reached. We also see that long
trajectories with small step sizes exhibit chaotic behavior.

All our data show a clear exponential instability in $||\Dd U||$ as a function
of $\trjlen$, so we fitted the data over the range of $\trjlen$ for which this
exponential behavior was obvious and extracted a characteristic exponent~$\nu$
(which we shall henceforth call the Liapunov exponent, although this might not
always be quite the correct terminology). In Figure~\ref{exponent-dt} we show
the Liapunov exponent as a function of the step size $\dt$ for pure $SU(3)$
gauge theory, and QCD with heavy and light dynamical fermions.
The results show the same qualitative behavior as that of our toy model which
was exhibited in Figure~\ref{fluc} on page~\pageref{fluc}, except that
\begin{itemize}
  \item The ``wall'' where the integration instability sets in is at a different
  value of $\dt$ (about $0.6$ for the pure gauge theory and heavy fermion cases,
  and about $0.1$ for the light fermion case). This probably just reflects the
  different highest frequencies of the systems.

  \item The exponent is not zero for small $\dt$, but has some fixed
  non-vanishing value. This is just the evidence for chaotic continuous time
  dynamics discussed earlier.
\end{itemize}
The integration instability is easily avoided by choosing a sufficiently small
step size, which is forced upon us anyhow as the lattice volume becomes large if
we want to maintain a reasonable Metropolis acceptance rate. This, of course, is
exactly what we mean by saying that the integration instability is a finite
volume effect. Furthermore, it is clear that even a $4^4$ lattice is
sufficiently large for us to be driven well away from the ``wall'' for all the
parameters we have considered.

We shall therefore turn our attention to the chaotic instability which is
present for any value of $\dt$, and in the next section we shall investigate the
dependence of $\nu$ upon the parameters of our lattice field theories.

\subsection{Parameter Dependence of Liapunov Exponent}
\label{liapunov-parameter-dependence}

\begin{figure}
  \epsfxsize=0.7\textwidth
  \centerline{\epsfbox{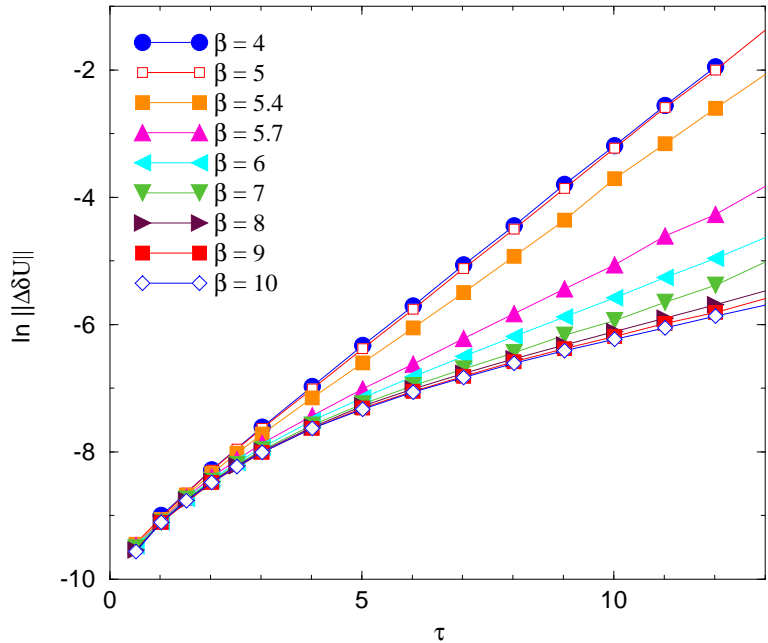}}
  \caption[beta-ddu-tau]{$||\Dd U||$ as a function of trajectory length
    $\trjlen$ for a variety of values of $\beta$ for pure $SU(3)$ gauge theory
    on $8^4$ lattices.}
  \label{beta-ddu-tau}
\end{figure}

\begin{figure}
  \epsfxsize=0.7\textwidth
  \centerline{\epsfbox{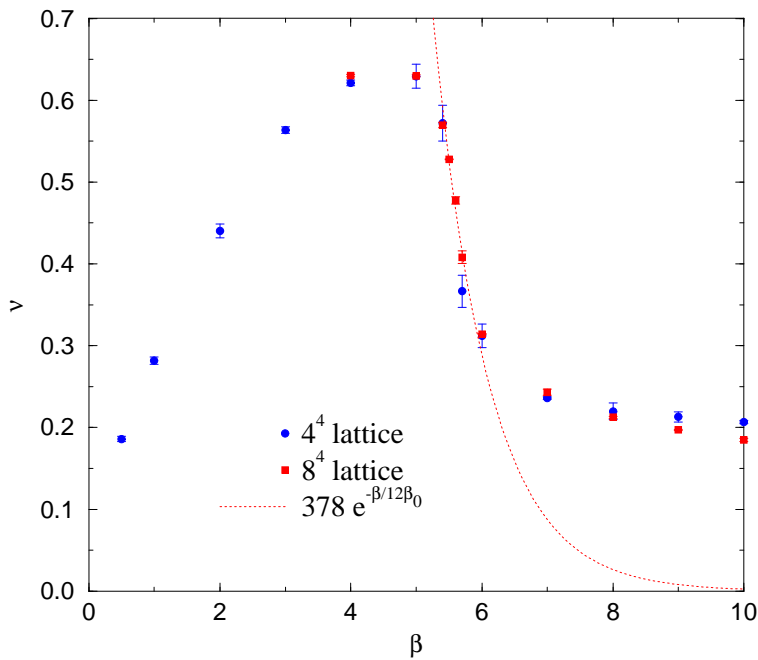}}
  \caption[liapunov-beta-fit]{The Liapunov exponent $\nu$ is shown as a function
    of $\beta$ for pure $SU(3)$ gauge theory on $4^4$ and $8^4$ lattices. The
    curve is a one-parameter fit of the functional form given in
    eq.~(\ref{asymptotic-scaling}) to four of the $8^4$ data points
    ($\beta=5.4$, $5.5$, $5.6$, and $5.7$). The data was measured on three $4^4$
    configurations and two $8^4$ configurations.}
  \label{liapunov-beta-fit}
\end{figure}

We investigated the dependence of the Liapunov exponent $\nu$ on the coupling
constant $\beta$ for pure $SU(3)$ gauge theory.\footnote{Somewhat similar
results have been reported by Jansen and Liu \cite{jansen96a}, although their
data does is not obviously consistent with ours, especially for large~$\beta$.}
The results are shown in Figure~\ref{beta-ddu-tau} and~\ref{liapunov-beta-fit}.
In the former we plot $\ln||\Dd U||$ against $\trjlen$ for a variety of values
of $\beta$. All of the curves exhibit a very clear exponential instability: for
small $\beta$ the relation is almost completely linear on this semi-logarithmic
plot, whereas for large $\beta$ there is some initial curvature before a  linear
region is reached. The slope (i.e., the Liapunov exponent) decreases very
rapidly between $\beta=5.4$ and $6.0$, which is just below the finite-size
analogue of the finite temperature deconfinement transition on a $8^4$ lattice.

In Figure~\ref{liapunov-beta-fit} we plot the measured values of the Liapunov
exponent $\nu$ as a function of $\beta$ for both $4^4$ and $8^4$ lattices. For
small $\beta$ the lattice theory is in the strong coupling regime and does not
obey the asymptotic scaling behavior mandated by the renormalization group
equations and the perturbative $\beta$-function. For large $\beta$ the system is
in a tiny box, and is thus in the deconfined phase. Therefore we can at best
only trust the data in some ``scaling window'' near $\beta=5.7$. In this region
we have fitted the $8^4$ data to our suggested asymptotic scaling form of
eq.~(\ref{asymptotic-scaling}) and we find the fit surprisingly good, especially
considering that there is only one free parameter.
It is dangerous to rely too much on asymptotic scaling results obtained on such
small lattices, so we will content ourselves with the statement that our data is
consistent with our hypothesis that $\nu\xi$ is a constant. The value of $\nu$
is also sufficiently small that the amplification of rounding errors for
trajectories of length $\trjlen=\xi$ is a factor of $O(1)$, and thus
unimportant.

We do not have a clear understanding of why the value of $\nu$ at large $\beta$
is so large for $4^4$ lattices, other than the obvious suggestion that this is a
finite size artefact. The data from $8^4$ lattices shows a small decrease in
$\nu$ for large $\beta$, but this is not convincing evidence that the data will
eventually fall onto our hypothetical asymptotic scaling curve.

\begin{figure}
  \epsfxsize=0.7\textwidth
  \centerline{\epsfbox{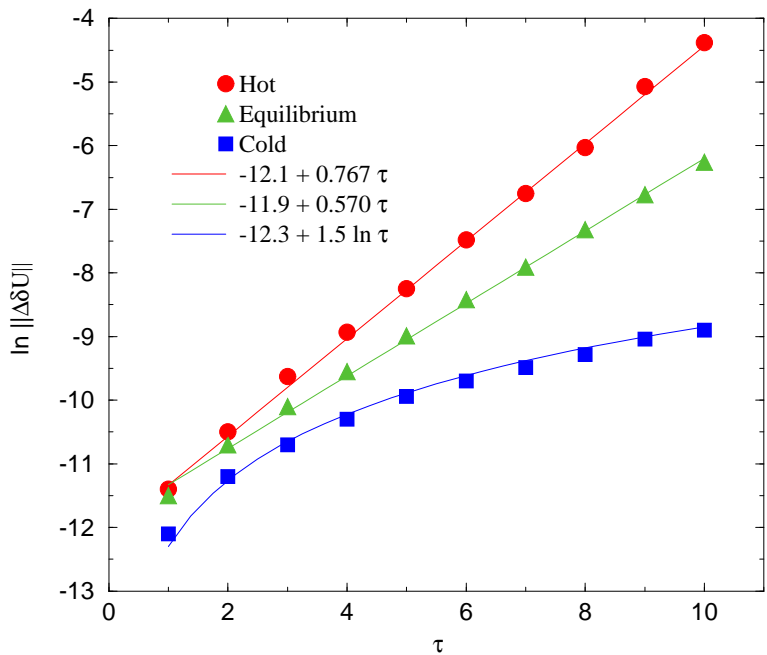}}
  \caption[config-order]{$\ln||\Dd U||$ is shown as a function of trajectory
    length $\trjlen$ (with $\beta=5.1$, $\kappa=0.16$, and $\dt=0.1$ on a $4^4$
    lattice) for starting gauge field configurations which are ``hot'' (all link
    variables $U_\mu(x)$ chosen independently according to $SU(3)$ Haar
    measure), ``equilibrium'' (a typical configuration chosen from the
    equilibrium ensemble), and ``cold'' (all link variables $U_\mu(x)=1$). Fits
    of the form $||\Dd U||=ke^{\nu\trjlen}$ for the hot and equilibrium
    configurations and $||\Dd U||=k\trjlen^{3/2}$ for the cold configuration are
    shown.}
  \label{config-order}
\end{figure}

We can gain a little more understanding of the mechanism by which $\nu$ could
decrease as we approach the continuum limit by noting that $\nu$ depends on
$\beta$ in two ways: there is an explicit $\beta$-dependence of the equations of
motion, and there is an implicit $\beta$-dependence in the equilibrium ensemble
of configurations over which $\nu$ is measured. This latter effect is
illustrated in Figure~\ref{config-order}, where we plot $\ln||\Dd U||$ against
$\trjlen$ for $\beta=5.1$ and $\kappa=0.16$ on a $4^4$ lattice for three
different configurations, one hot, one cold, and one chosen at random from the
equilibrium distribution.\footnote{We verified that the results for the
equilibrated configuration did not change significantly when we selected another
such configuration.} The hot configuration yields a larger Liapunov exponent
than the equilibrium one, and the cold configuration is consistent with a
power-law dependence of $||\Dd U||$ on~$\trjlen$. This result suggests that we
should not expect to get the correct continuum result from lattices which are
too small to be in the confined phase (even if we wanted to consider a finite
temperature system in the deconfined phase we should at least ensure that the
spatial volume is large enough).

In Figure~\ref{config-order} we also show two-parameter fits to the hot and
equilibrium trajectories of the form $||\Dd U||\propto e^{\nu\trjlen}$, and a
one-parameter fit to the cold trajectory of the form $||\Dd U||\propto
\trjlen^{3/2}$. These fits are stable if we add more free parameters by fitting
to a combination of an exponential and power dependence. The power-law behavior
of the cold trajectory may be explained by a combination of a factor of
$\sqrt\trjlen$ for the random walk evolution of rounding errors and a linear
factor characteristic of the divergence of nearby trajectories for interacting
dynamical systems in the stable (non-chaotic) region of phase space.
\note[Test]{This could be checked by varying $\dt$}.

\subsection{Reversibility and Conjugate Gradient Accuracy}

Our final topic is to investigate the effect of inaccurate CG solutions on
instabilities and reversibility for dynamical fermion computations.

\begin{figure}
  \epsfxsize=0.7\textwidth
  \centerline{\epsfbox{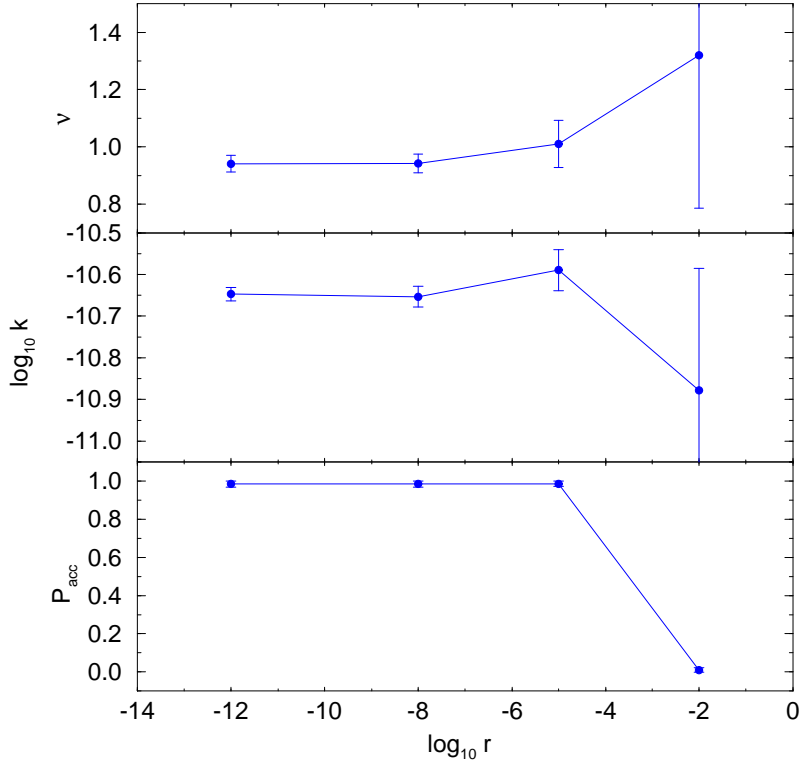}}
  \caption[residual]{The dependence of $\pacc$ and $||\Dd U||$ on the CG
    residual $r$. We write $||\Dd U|| = ke^{\nu\tau}$, where the  coefficient
    $k$ and exponent $\nu$ are functions of $r$. A time-symmetric zero initial
    CG vector was used. The data is for $4^4$ lattices at $\beta=3.5$ and
    $\kappa=0.2225$ with $\dt=0.01$. \note[More data for residual dependence]{It
    would be nice to have some more data between $r=10^{-2}$ and $r=10^{-5}$.}}
  \label{residual}
\end{figure}

In Figure~\ref{residual} we show the dependence of the Liapunov exponent $\nu$
and the Metropolis acceptance rate as a function of the logarithm of the CG
residual for a time-symmetric (zero) initial CG vector. We define the CG
residual for the approximate solution $x'$ of the linear equations $Ax=b$ as
$r\equiv||b - Ax'||/||b||$. We also show the logarithm of the coefficient $k(r)$
for completeness.
The results clearly show that the choice of residual has no effect for
$r<10^{-5}$, and that for $r>10^{-2}$ the acceptance rate is essentially zero.
There would seem to be no benefit from finding the solution more accurately than
is needed to give a reasonable acceptance rate.

\begin{figure}
  \epsfxsize=0.7\textwidth
  \centerline{\epsfbox{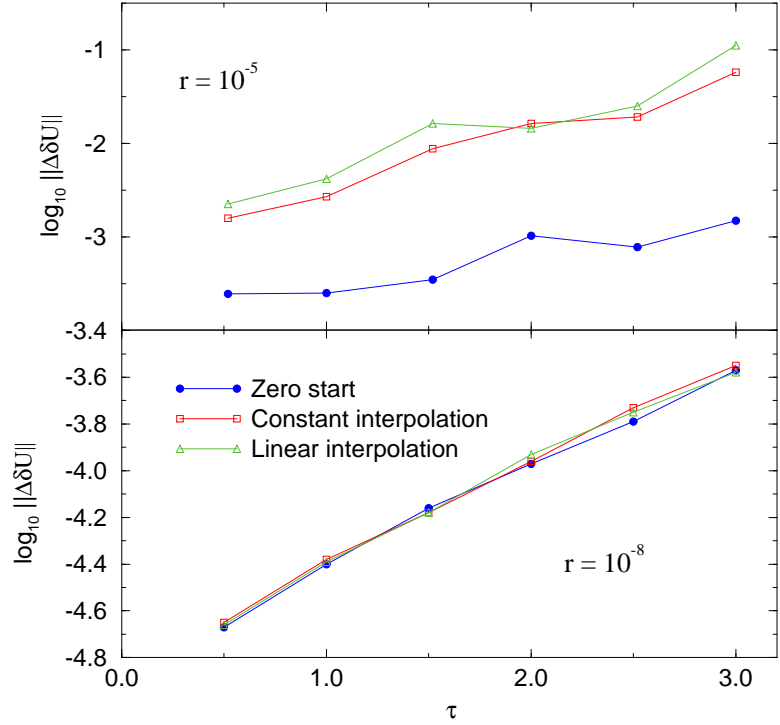}}
  \caption[interpolate]{The dependence of $\ln||\Dd U||$ on $\trjlen$ for
    different choices of initial vector for the CG algorithm. The results  for
    the time-symmetric zero start are to be compared with those for which the
    initial CG vector was chosen to be the solution from the previous time step
    (constant interpolation), and a linear extrapolation from the previous two
    time steps (linear interpolation). For a residual $r=10^{-8}$ the initial
    vector has little effect upon the solution obtained, but for the larger
    residual $r=10^{-5}$ the explicit violation of reversibility is very
    significant. The data is from $4^4$ lattices with $\beta=3.5$,
    $\kappa=0.2225$, and $\dt=0.02$.}
  \label{interpolate}
\end{figure}

The effect of improving the convergence of the CG algorithm by using a
time-asymmetric initial vector is illustrated in Figure~\ref{interpolate}. For a
residual $r=10^{-8}$ the solution vector is sufficiently independent of the
starting vector that the choice of initial vector has no effect on the
reversibility (or otherwise) of the trajectory. On the other hand, for
$r=10^{-5}$ there is a large difference between the time-symmetric start and the
intrinsically irreversible time-asymmetric starts.

It is clear that the reduction in the number of CG iterations required to reach
a prescribed residual obtained by using an starting vector derived from previous
nearby solutions \cite{brower95a} needs to be offset against the need to have a
more accurate solution in order to avoid increased reversibility errors. To
reach definitive conclusions more extensive studies would have to be done.

\section{Conclusions}
\begin{itemize}
  \item Instabilities and the concomitant amplification of rounding errors
  caused by inaccurate numerical integration of the equations of motion should
  not be a problem if the integration step size is chosen suitably.

  \item The instabilities due to integration errors become unimportant as
  $V\to\infty$.

  \item We hypothesize that the Liapunov exponent falls as $\nu\propto\xi^{-1}$
  for sufficiently large correlation lengths $\xi$, and therefore we can choose
  trajectories of length $\trjlen\propto\xi$ which may reduce critical slowing
  down without suffering from exponentially large amplification of rounding
  errors.

  \item The measured values of the Liapunov exponent are of $O(1)$, so the
  exponential amplification factor for trajectories of length
  $\trjlen\approx\xi$ are this only $e^{\nu\xi}=O(1)$.

  \item We have found examples with large reversibility errors but reasonable
  acceptance rate and small $|\Dd H|$. It is always prudent to verify that
  $||\Dd U||$ is small.
\end{itemize}

\section*{Acknowledgements}
We would like to thank Karl Jansen for kindly sending us a draft of
Ref.~\cite{jansen96a}.

This research was supported by by the U.S. Department of Energy through Contract
Nos. DE-FG05-92ER40742 and DE-FC05-85ER250000.

\bibliographystyle{scri-unsrt}
\bibliography{adk,lattice-bibliography}

\begin{thebibliography}{10}

\bibitem{kennedy87a}
A.~D. Kennedy.
\newblock {Hybrid} {Monte} {Carlo}.
\newblock In A.~Billoire, R.~Lacaze, A.~Morel, O.~Napoly, and J.~Zinn-Justin,
  editors, {\em Field Theory on the Lattice}, volume~B4 of {\em Nuclear Physics
  (Proceedings Supplements)}, 1988.
\newblock 1987 International Conference on Field Theory on a Lattice,
  {Seillac}, {France}.

\bibitem{duane87a}
Simon Duane, A.~D. Kennedy, Brian~J. Pendleton, and Duncan Roweth.
\newblock {Hybrid} {Monte} {Carlo}.
\newblock {\em Phys. Lett.}, 195B(2):216--222, 1987.

\bibitem{kennedy90a}
A.~D. Kennedy.
\newblock The theory of {Hybrid} {Stochastic} algorithms.
\newblock In P.~H. Damgaard et~al., editors, {\em Probabilistic Methods in
  {Quantum} Field Theory and {Quantum} {Gravity}}, pages 209--223, New York,
  1990. NATO, Plenum Press.
\newblock Lectures given at the workshop on ``Probabilistic Methods in
  {Quantum} Field Theory and {Quantum} {Gravity},'' {Carg\'ese}, August 1989.

\bibitem{horowitz90a}
Alan~M. Horowitz.
\newblock A generalized guided {Monte} {Carlo} algorithm.
\newblock {\em Phys. Lett.}, B268:247--252, 1991.

\bibitem{gausterer89a}
H.~Gausterer and M.~Salmhofer.
\newblock Remarks on global {Monte} {Carlo} algorithms.
\newblock {\em Phys. Rev.}, D40(8):2723--2726, October 1989.

\bibitem{gupta90a}
Sourendu Gupta, Anders {Irb\"ack}, Frithjof Karsch, and Bengt Petersson.
\newblock The acceptance probability in the {Hybrid} {Monte} {Carlo} method.
\newblock {\em Phys. Lett.}, B242:437--443, 1990.

\bibitem{kennedy91b}
A.~D. Kennedy and Brian~J. Pendleton.
\newblock Acceptances and autocorrelations in {Hybrid} {Monte} {Carlo}.
\newblock In Urs~M. Heller, A.~D. Kennedy, and Sergiu Sanielevici, editors,
  {\em Lattice '90}, volume B20 of {\em Nuclear Physics (Proceedings
  Supplements)}, pages 118--121, 1991.
\newblock Talk presented at ``Lattice '90,'' Tallahassee.

\bibitem{kennedy91a}
A.~D. Kennedy and Brian~J. Pendleton.
\newblock Some exact results for {Hybrid} {Monte} {Carlo}.
\newblock In preparation, 1996.

\bibitem{kennedy95a}
A.~D. Kennedy, Robert~G. Edwards, Hidetoshi Mino, and Brian~J. Pendleton.
\newblock Tuning the generalized {Hybrid} {Monte} {Carlo} algorithm.
\newblock In Tien~D. Kieu, Bruce H.~J. McKellar, and Anthony~J. Guttmann,
  editors, {\em Lattice '95}, volume B47 of {\em Nuclear Physics (Proceedings
  Supplements)}, pages 781--784, 1995.
\newblock Proceedings of the 13th International Symposium on Lattice Field
  Theory, {Melbourne}, {Australia}, 11--15 July 1995.

\bibitem{edwards92a}
Khalil~M. Bitar, Robert~G. Edwards, Urs~M. Heller, and A.~D. Kennedy.
\newblock On the dynamics of light quarks in {QCD}.
\newblock In Jan Smit and Pierre van Baal, editors, {\em Lattice '92}, volume
  B30 of {\em Nuclear Physics (Proceedings Supplements)}, pages 249--252, March
  1993.
\newblock Proceedings of the International Symposium on Lattice Field Theory,
  {Amsterdam}, the {Netherlands}, 15--19 {September} 1992.

\bibitem{jansen95a}
Karl Jansen and Chuan Liu.
\newblock {Kramer's} equation algorithm for simulations of {QCD} with two
  flavors of {Wilson} fermions and gauge group {$SU(2)$}.
\newblock {\em Nucl. Phys.}, B453:375--394, 1995.

\bibitem{jansen96a}
Karl Jansen and Chuan Liu.
\newblock {Liapunov} exponents and the reversibility of molecular dynamics
  algorithms.
\newblock Preprint, DESY, 1996.

\bibitem{nielsen96a}
Holger~B. Nielsen, H.~H. Rugh, and S.~E. Rugh.
\newblock Chaos and scaling in classical non-abelian gauge fields.
\newblock Chao--dyn/9605013, Niels Bohr Institute, May 1996.

\bibitem{brower95a}
Richard~C. Brower, T.~Ivanenko, A.~R. Levi, and K.~N. Orginos.
\newblock Chronological inversion method for the dirac matrix in {Hybrid}
  {Monte} {Carlo}.
\newblock Preprint, Boston University, 1995.

\end{thebibliography}
\end{document}
\bye